# Structural, micro-structural/morphological and magnetic properties of RECrO$_3$ (RE = La, Nd, Sm, Eu and Gd) Orthocromites

Anurag Yadav and Anuj Kumar*

*Department of Physics and Materials Science Engineering, Jaypee Institute of Information Technology, Noida-201304, Gautam Budda Nagar, Uttar Pradesh, India*

## Abstract

We report here the structural, micro-structural/morphological and detailed magnetic properties of RECrO$_3$ (RE = La, Nd, Sm, Eu and Gd) Orthocromites. A series of RECrO$_3$ (RE = La, Nd, Sm, Eu and Gd) Orthocromites polycrystalline samples with uniform particle size were synthesized via standard solid-state reaction method. LaCrO$_3$ (LCO), NdCrO$_3$ (NCO), SmCrO$_3$(SCO), and EuCrO$_3$ (ECO) crystallizes in single phase of orthorhombic unit cell structure with *Pnma (No. 62)* space group while GdCrO$_3$ (GCO) found to crystallize in orthorhombic phase with *Pbnm* space group. SEM images confirms the systematic growth of all studied samples Grain growth takes place during the sintering process and images shows the polyhedral shape with varying average grain size ranging from approximately 1.0μm to 3μm confirm the bulk synthesis of studied RECrO$_3$. Lowest average grain size (~ 1.1μm) was observed for LaCrO$_3$ and maximum (~ 2.8μm) for EuCrO$_3$. The shapes of the grain size changes because of crystal lattice distortion induced by various rare earths (La, Nd, Sm, Eu and Gd) at A site. Energy dispersive X-ray Analysis (EDAX) reveals the elements confirmation in different studied RECrO$_3$ samples. Within the detection limit of EDAX no impurity elements were observed in different as synthesized RECrO$_3$ samples. Temperature dependent DC magnetization (M-T) studies showed an antiferromagnetic transition (T$_N$) attributed to Cr$^{3+}$ spins in all samples at different temperature. At low temperature spin-reorientation (T$_{SR}$) of rare earth ion was also observed in NdCrO$_3$, SmCrO$_3$ at around 50K and the same is observed at around 30K in EuCrO$_3$ and the same was missing in GdCrO$_3$ because of large moment of Gd$^{3+}$ ions attributed by ground state spectroscopic term $^8S_{7/2}$ by external applied magnetic field. The magnetization versus applied magnetic field (M-H) loops exhibited ferromagnetic like behavior at 10K and 100K and paramagnetic at 300K for GdCrO$_3$.

# 1. Introduction

Now a days utmost attempts have been made for development of novel materials having improved functional properties for use in variety of applications as storage and conversion of energy, sensing technologies and chemical engineering [1]. In view of this context, synthesis and study of rare-earth perovskite orthochromites (RECrO$_3$), RE denotes rare earth element, is gaining much attention because of their striking magnetic properties and possible application of device fabrication and fast spin dynamics. Perovskite materials are the most useful and diverse materials in realm of solid state due to variations in their composition and structure. The term "perovskite" termed as the mineral "CaTiO$_3$" and it was discovered by Gustav Rose (1839) and Victor Goldschmidt studied its crystal structure for the first time. Perovskite materials exhibit different forms that may include ABO$_3$-perovskite (BaTiO$_3$ or CaTiO$_3$), A$_2$BO$_4$-layered perovskite (Sr$_2$RuO$_4$ or K$_2$NiF$_4$), A$_2$BB'O$_6$-double perovskite (Ba$_2$TiRuO$_6$) and A$_2$A'B$_2$B'O$_9$-triple perovskite (La$_2$SrCo$_2$FeO$_9$) etc. [2, 3]. Since perovskites have remarkable electrical, magnetic as well as catalytic properties [4, 5] this makes them efficient to use [6] for numerous applications such as dielectrics, solar cells, quantum computing, and arising technologies like spintronic [7-14]. Rare earth orthochromites (RECrO$_3$) belong to perovskite family, wherein B-site is occupied by Cr$^{3+}$ ions while A-site is occupied by any one of the rare earth ions as RE = La, Pr, Nd, Sm, Eu, Gd, Tb, Dy, Tm, Ho, Er, Ce and Yb [15-18]. In RECrO$_3$ the weak ferromagnetism emerges at low temperature due to the interactions of magnetic ordering of 4f electrons of RE ions with 3d electrons of Cr$^{3+}$ ions. Cr$^{3+}$ spins also create an effective magnetic field acting of RE sublattice. Though, the precise origin of weak magnetization and produced magnetic field at low temperature is known. Some research group claims that it is induced by the weak magnetization of Cr sublattice, while others purposed that it involves all type of possible magnetic ordering (ferromagnetism, and G, A, C-type anti-ferromagnetism) of the same Cr sub-lattice. The competition between the magnetic ordering of RE$^{3+}$ and Cr$^{3+}$ leads to the various kind of magnetism dominated phenomena like temperature and field induced time reversal magnetization, exchange bias and spin reorientation [19, 20]. Magnetism dominated properties are used in various industrial applications like random access memory, thermo-magnetic switches and thermally assisted magnetic heads [21, 22]. G-type canted anti-ferromagnetism also reported in RECrO$_3$ below the Neel temperature (T$_N$) in the range of 120-290K range, due to the ionic size of rare earth ion. Weak ferromagnetism and Dzyaloshinskii-Moriya (DM) interaction is the origin of canted anti-ferromagnetism in

RECrO3 system [23]. In last two decades the lots of efforts to put by materials community to obtain new magnetic refrigeration technology and Orthocromites is also a potential candidate to work as a material for low temperature magnetic refrigeration technique [24, 25].Owing to these characteristic features, these materials find applications as gas sensors [26], catalysts and magneto-optical devices [27]. Since these materials also possess remarkable optical and magnetic properties these materials make them effective for use in light emitting diodes (LED) [28], optoelectronics [29], refrigeration [30], spintronics devices [31] and thermally assisted magnetic random-access memory (MRAM) [32]. Magnetic properties of RECrO3 at variable temperature ranges attributed to three magnetic interactions namely $Cr^{3+}$– $Cr^{3+}$, $Cr^{3+}$–$RE^{3+}$, and $RE^{3+}$- $RE^{3+}$ interactions. These types of phenomena responsible for spin reorientation [32], negative magnetization [33] and exchange bias effect [32]. Structure of RECrO$_3$ materials at room temperature possess space groups i.e., *Pbnm* or *Pnma* groups and properties of these groups got affected due to structural distortion. Various parameters affecting structure of these groups may include t-e hybridization [34], oxygen stoichiometry and nature of doping element and ionic size of rare earth elements $RE^{3+}$ [32].

Chromite of lanthanum (LaCrO$_3$, LCO) is a member of RECrO$_3$ that is refractory oxide [19, 35] and is a p-type semiconductor having a bandgap of 3.4 eV [32]. Several methods for synthesis of LCO include sol–gel [20], microwave sintering, hydrothermal [36], combustion synthesis [21] and conventional solid-state reaction [32]. These materials draw researcher's attention owing to their remarkable chemical stability, efficient catalytic performance [32], infrared emissivity [22] and good electrical conductivity [32]. These characteristic properties of LCO make them efficient for use as solar absorber materials [23], coatings [37], catalysts [21] and solid oxide fuel cell interconnects [32]. In addition to LCO, chromite of samarium (SmCrO$_3$, SCO) is of great interest in today's society. These materials are paramagnetic to canted AFM transition of nearly 192K that may be attributed to the ordering of the spins of $Cr^{3+}$. These materials also possess spin reorientation transition (at low temperature of nearly 34K) that are mainly due to the rotation of $Cr^{3+}$ spins taking place from one direction to the other [24, 38]. When it comes to the chromite of neodymium (NdCrO$_3$, NCO), it is observed that magnetization curve of NCO differs to a greater extent than those typically found in the RECrO$_3$ [25]. NCO in general possess remarkable ferroelectric property that might be due to the structural transformation from the centrosymmetric space group *Pnma* to a non-centrosymmetric space group Pna2$_1$ [39]. Chromite of gadolinium (GdCrO$_3$, GCO) is one of the promising orthochromites. This chromite possesses magnetic rare earth ion having an orthorhombic crystal structure along with *Pbnm* space group. Owing to spin canting, there is an antiparallel alignment of $Gd^{3+}$ moment with that of weak ferromagnetic part of $Cr^{3+}$ ions leading to the presence of remarkable MR behavior in it [40]. Further, chromites of europium ($EuCrO_3$, ECO) have orthorhombic distorted perovskite structure. In the crystallized form, it has four formula units per unit cell. Overall, in $RECrO_3$, the magnetic ordering of *3d* spins of transition metal and 4f moments rare earth ions, respectively, are accountable for magneto-electric coupling via lattice distortion. These materials are widely employed in numerous fields including ferroelectric, multilayer ceramic capacitors and magnetic field sensors [41].

Numerous works have been reported on the potential of perovskite materials as Song et al., worked on the development of the novel multiferroic-like nanocomposite using the high-pressure torsion (HPT) method. The composite, consisting of exchange-interacting ferroelectric (FE) and ferromagnetic (FM) phases, shows significantly reduced particle sizes and improved magnetic and electrical properties even at the room temperature. Furthermore, the study reveals an exceptional increase in residual polarization with high pressure, suggesting the nanocomposite's potential for use in ultra-high-pressure monitoring devices and high-speed computing circuits with low switching energy [42]. In another study [43], a magnetoactive elastomer (MAE) with $La_{0.6}Ag_{0.2}Mn_{1.2}O_3$ nanoparticles was prepared and investigated that exhibits the superparamagnetic behavior and low coercivity at room temperature. This study involved the coupling reaction between magnetization and magnetoelasticity showing the critical features near the Curie temperature ($T_C$ = 308 K). Above $T_C$, the magnetoelastic properties disappear, while below this temperature, they can be restored that make these materials as the smart materials that find extensive uses in devices with self-regulating magnetoelastic properties even at room temperature. Similarly, the magnetic and the electrocatalytic properties of the cobaltite ($La_{0.6}Sr_{0.4}CoO_3$ (LSCO)) and $La_{0.6}Sr_{0.3}Mn_{1.1}O_3$ (LSMO) [44] also been reported.

Keeping in view the above discussion, from the literature [45-52], it reveals that a scare combined study of structural, magnetic and electrical transport properties on the $RECrO_3$ (RE = La, Nd, Sm, Eu & Gd). Though some fascinating studies of structural, magnetic and electrical transport properties on individual (both pure and doped) $RECrO_3$ (RE = La, Sm, Nd, Gd, Eu) are reported but still there is a need to study their properties in a more comprehensive and combined manner. Therefore, a combined study on the structural, micro-structural/morphological, and magnetic properties of the $RECrO_3$ (RE = La, Nd, Sm, Eu & Gd) are reported.

The present study is novel in a sense that it involves the investigation of the comparative analysis of structural, micro-structural as well as the magnetic properties of a series of RECrO$_3$ (where RE = La, Nd, Sm, Eu, and Gd) orthochromites synthesized with the help of a solid-state reaction method. This study provides a comprehensive analysis of the studied materials with key focus on the effects of different rare earth elements on grain size, phase formation in addition to the magnetic transitions. Additionally, this work involved the detailed structural analysis combined with magnetic property investigations at variable temperatures, offering new insights into spin reorientation phenomena as well as the potential applications in magnetic refrigeration and spintronics. The study of temperature-dependent studies of magnetic transitions and the explanation of the relationship between the structural distortions as well as in the magnetic characteristics make this study different from previous works in the field.

## 2. Experimental details

All polycrystalline bulk materials of RECrO$_3$ (RE = La, Nd, Sm, Eu and Gd) were synthesized through standard solid-state reaction route from stoichiometric ratio of 99.99% Rare Earth (RE) metal oxides (RE$_2$O$_3$) and the transition metal oxide (Cr$_2$O$_3$). Chemical name, purity (%) and brand name of all chemicals are La$_2$O$_3$ 99.99% Sigma Aldrich, Nd$_2$O$_3$ 99.9% Alfa Aesar, Sm$_2$O$_3$ 99.9% Sigma Aldrich, Eu$_2$O$_3$ 99.9% Sigma Aldrich, Gd$_2$O$_3$ 99.9% Sigma Aldrich, and Cr$_2$O$_3$ 99.0% Sigma Aldrich. After mixing pure chemicals, we get a compound like RECrO$_3$ (RE = La, Nd, Sm, Eu & Gd). These two powders were ground together in an agate mortar and heat treated at 1100°C and 1250°C each for 24h at the rate of 5°C/minute with proper intermediate grindings. After the 1250°C calcination the final powder pressed in circular shaped pellets. The pressure applied during the compacting of the materials was 5-6 tons, and the size of the pellets (circular tablet) is around 13 mm in diameter and 1.5 mm in thickness. These circular shaped pellets again sintered at 1400°C for 12h hold time and subsequently cooled down to room temperature through natural cooling to get the final samples. X-ray diffraction (XRD) of all samples were performed at room temperature within the range of 20°-80° of 2θ value with the step size 0.02° using the *Rigaku* made X-ray diffractometer with Cu-K$_\alpha$ line radiation (λ = 1.54Å) for the structural analysis. Rietveld analysis of the XRD data of each sample was done using standard *FullProf* program which is free available on website. For microstructure analysis or surface morphology of the samples we performed the scanning electron microscope (SEM) characterization and the corresponding EDX/EDS elemental analysis of each freshly fractured surface of samples were taken using FESEM JEOL JSM 7600 Plus. To probe the magnetic properties, temperature and field dependent magnetization (M-T and M-H) were measured on standard Physical Property Measurement System (PPMS-14T Cryogenics, United Kingdom).

## 3. Results and discussion

The quality of the reported materials (RECrO$_3$) is very-very important for any meaningful discussion particularly in materials science/experimental condensed matter physics because a minute foreign phase or impurity (unreacted rare earth oxides/transition metal oxides can ruin magnetic properties of the studied material. During synthesis we have gone through various calcination's steps and then final sintering just to avoid the development of any foreign phase and to get the phase pure RECrO$_3$. No foreign phase or impurities were observed in all as synthesized RECrO$_3$ (RE = La, Nd, Sm, Eu, & Gd) samples as far the detection limit of XRD. Structural analysis was performed for each sample by using the *FullProf* program that conformed all samples crystallize in single phase with different unit cell structure and space group. **Figure 1 (a-e)** depicts the Rietveld fitted XRD pattern of all samples viz; LaCrO$_3$ (LCO), NdCrO$_3$ (NCO), SmCrO$_3$ (SCO), EuCrO$_3$ (ECO) and GdCrO$_3$ (GCO). Best refinement was obtained when LCO, NCO, SCO and ECO fitted in *Pnma* space group with orthorhombic unit structure while GCO found to be in again orthorhombic phase but with different space group *Pbnm*. All Rietveld refined parameters (lattice parameters, cell volume, Wyckoff position, site occupancy and $\chi^2$ value) of all compounds are shown in **Table 1 and 2.** It is observed that lattice parameters and cell volume of all samples are found comparable to the reported data [53]. Volume of the ECO and GCO is smaller than the volume of LCO because of the larger ionic size of La$^{3+}$ ($r_{La}^{3+}$ = 1.216Å ) as comparable to Eu$^{3+}$/Gd$^{3+}$while not much appreciable change was observed in the volume of NCO and SCO. As far as stability of the perovskite structure is concerned, tolerance factor (*t*) plays an important role to decide the same. The tolerance factor is defined by $t = (R_A + R_O)/\sqrt{2}(R_B + R_O)$, where R$_A$, R$_B$, and R$_O$ are the average ionic radii of rare-earth cation (RE$^{3+}$), chromium cation (Cr$^{3+}$) and oxygen anions (O$^{2-}$) respectively. The calculated values of tolerance factor (t) for all studied samples are in good agreement with the reported standard values 0.75 < t < 0.90 for the distorted orthorhombic structure. For further analyzing the orthorhombic distortion from a perfect cubic structure, one more parameter called the orthorhombic strain factor S = $2(a - c)/(a + c)$ was calculated for all studied samples. The tolerance factor (t) and orthorhombic lattice strain (S) both are calculated for all studied samples and shown in **Table 1.** It was also observed that as tolerance factor (t) decreases as the orthorhombic lattice strain increases which leads to the orthorhombic distortion.

To check the surface morphology/micro-structure we performed the Scanning Electron Microscopy (SEM) and Electron Dispersive X-ray of LCO, NCO, SCO, ECO and GCO samples as shown in **figure 2 (a-f)**. The systematic grain growth takes place during the sintering process for all the samples and SEM images depicts the polyhedral shape grains with variable grain sizes. Grains seem well connected in LCO and NCO samples and some porosity were observed in SCO and GCO samples while in case of ECO grains seems interleaved in some areas. The average grain size of all the studied samples was calculated by using the Image J software. The average grain size estimated for all studied samples is varied in the range of 1-3μm. The average grain size for LCO, NCO, SCO, ECO and GCO are 1.1μm, 1.5μm, 1.8μm, 2.8μm and 2.3μm respectively. The shapes of the grain size changes because of crystal lattice distortion induced by various rare earths (La, Nd, Sm, Eu and Gd) at A site. Energy dispersive X-ray Analysis (EDAX) reveals the elements confirmation in different studied $RECrO_3$ samples and it was confirmed that all samples are approximately in intended ratio as they were mixed and no foreign elements except carbon on the surface was detected in EDAX analysis. Based on EDAX data the atomic/weight % of the elements for all studied samples are shown in **Table 3**.

Due to different size of rare earth ions ($RE^{3+}$) and $Cr^{3+}$ ions these orthocromites are not perfect cubic in structure. The $CrO_6$ octahedra distorted the O-Cr-O angles other than 180º and reduce the anti-ferromagnetic double exchange interaction in neighbors of $Cr^{3+}$ ions [54]. Ortho-chromites show G-type anti-ferromagnetic ordering of $Cr^{3+}$ ions at Neel temperature [55] but due to the slight canting in $CrO_6$ octahedral a weak ferromagnetic behavior also observed [56]. Below the room temperature all samples shows paramagnetic to anti-ferromagnetic transition followed the Neel temperature and spin reorientation temperature in the temperature range of 240-150K. **Figure 3(a-d)** depicts temperature dependent DC magnetization (M-T) of phase pure as synthesized $RECrO_3$ (RE = Nd, Sm, Eu & Gd) measured at applied field 1000 Oe and temperature range 5-300K with zero-field-cooled (ZFC) and field cooled (FC) protocol. The observed dc magnetization has relatively small value in ZFC mode as compared to FC mode. In **figure 3(a)** the spin ordering of $Cr^{3+}$ ions take place for NCO at around ($T_N$ = 230.5 K) which is the first magnetic ordering transition called the Neel temperature [57]. A wide λ-shaped peak was also observed in M-T measurement at $T_N$ due to the long-range cooperative ordering of Cr sublattice. In NCO $Cr^{3+}$ spin system discontinuously changes from high temperature $\Gamma_2$ ($F_x$, $C_y$, $G_z$) configuration to low temperature $\Gamma_1$ ($A_x$, $G_y$, $C_z$) at first spin orientation $T_{SR1}$ temperature which confirms that the Cr spin system is unaffected Cr-Nd interaction in NCO [58]. At low temperature there are two more transitions ($T_{SR1}$ = 40K and $T_{SR2}$ = 33K) was observed in NCO. The observation of $T_{SR1}$ at low temperature in NCO can only be explained due the distorted crystal structure. So, in case of NCO below $T_{SR1}$ Nd-Cr interaction in unaffected while above $T_{SR1}$ it is extremely strong. Field depended dc magnetization (M-H) of NCO is shown in **figure 4(a)** at different temperatures (5, 100 and 300K) in the applied field range of -30kOe to +30kOe. From M-H curves it is clear that no hysteresis was observed for NCO even at 10K. At 300K the behavior of M-H curve is like paramagnetic in the field range of -30kOe to +30kOe, and the similar behavior persists at lower temperatures (100K and 10K) though the value of magnetization increases as the temperature decreased. M-H curve at 10K have large magnetic moments as compared to the M-H observed at 100K and 300K which is attributed to the low temperature magnetic behavior of Nd. None of the M-H curves observed at different temperatures (10, 100 and 300K) shows saturation upto the applied magnetic field of 30kOe. However, there is a report [59] where the unsymmetrical small hysteresis behavior in NCO was observed and they believe this is due to the exchange bias, but no other reports confirmed their results. **Figure 3(b)** demonstrates the temperature dependent dc magnetization (M-T) of SCO under zero-field and field-cooled conditions with an applied magnetic field of 1000 Oe. In SCO the first magnetic ordering temperature observed as antiferromagnetic ordering of $Cr^{3+}$ spins ($T_N$ = 201K) with weak ferromagnetism. After settled down the $Cr^{3+}$ spins antiferromagnetically the magnetic moment which is the magnitude of torque exert in the presence of magnetic field, continues increases because of the weak ferromagnetism dominated by $Sm^{3+}$ ions in low spin state. On further lowering the temperature the spin reorientation of Cr-spins takes place at around below 40.5 K which is due to the change in Cr-spins configuration from $\Gamma_4$ ($G_x$, $A_y$, $F_z$) to $\Gamma_2$ ($F_x$, $C_y$, $G_z$). Origin of spin reorientation of $Cr^{3+}$ in orthocromites/orthoferrites is due to the anisotropic magnetic interactions between the rare earth $Sm^{3+}$ ions with $Cr^{3+}/Fe^{3+}$ ions at low temperature [60, 61]. So, SCO shows successive SR-FM-AFM transitions as we cool the sample in 1000 Oe applied field and almost same type of behavior PM-FM-AFM was observed in SmCoPO [62]. **Figure 4(b)** shows the magnetic field dependent dc magnetization (M-H) of SCO sample at temperatures (10, 100 and 300) with an applied field range of -30kOe to +30kOe. The behavior of SCO seems paramagnetic and no hysteresis loop was observed at 300K and 10K while SCO shows ferromagnetic like behavior and hysteresis loop was observed at 100K in the presence of same applied field range. The

Coercivity ($H_c$) and remanence magnetization ($M_r$) is 11.91 kOe and 0.30 emu/g respectively for the M-H loop observed at 100K. All M-H loops measured at different temperatures (10, 100 and 300K) are unsaturated upto the applied field of 30kOe while the value of magnetization increases as the temperature decreases. Such type of behavior may be due to the magnetic exchange of $Sm^{3+}$ with $Cr^{3+}$ ions in SCO.

**Figure 3(c)** demonstrators the temperature dependence magnetization behavior of ECO sample in the temperature range of 10-300K with applied magnetic field of 1000 Oe. ECO shows a weak spontaneous magnetization below the Neel temperature ($T_N$ = 177K) which is attributed due to the canted $Cr^{3+}$ moments which are antiferromagnetically aligned. Below the Neel temperature ECO behavior is like Pauli's paramagnetic as the magnetic moments continuously increases as lowering the temperature. On further lowering the temperature a second kind of transition was also observed in ECO at around 24K which is attributed to the spin reorientation of $Cr^{3+}$ spins. **Figure 4(c)** depicts the field dependent dc magnetization of ECO sample at temperature 10, 100 and 300K in the applied field range of -30kOe to +30kOe. The M-H at 100K and 300K shows paramagnetic like behavior while the M-H at 10K different and the S type shape in with rescannable coercive field ($H_c$) and remnant magnetization ($M_r$) which are the characteristics features of spin-glass behavior. At 10K and in low field regime the opening of M-H loop resembles the ferromagnetic like behavior of ECO. **Figure 3(d)** shows the dc magnetization (M-T) of GCO sample in zero-field cooled and field cooled situation measured at 1000 Oe in the temperature range of 10-300K. The branching of ZFC and FC starts at around ($T_N$ = 191K) which is the Neel temperature where $Cr^{3+}$ spins order antiferromagnetically in GCO. Below the Neel temperature the magnetization continuously increases as lowering the temperature which is due to the magnetic contribution of $Gd^{3+}$ ions. **Figure 4(d)** shows the typical isothermal magnetization of GCO sample at various temperatures (10, 100 and 300K) for applied field range of -30kOe to +30kOe. M-H loop at 300 K shows like a paramagnetic behavior while M-H loop at 10 and 100K shows sufficient opening and ferromagnetic behavior. For 10K M-H loop the coactivity ($H_c$) and remanence magnetization ($M_r$) is 10.25 kOe and 0.73 emu/g respectively while the same is observed for 100K M-H loop is 10.33 kOe and 0.69 emu/g respectively. The opening of M-H loop at low temperatures attributed to the magnetic nature of $Gd^{3+}$ ions at low temperature. Like others Orthocromites, the saturation magnetization was not observed in GCO sample. On lowering the temperature, the magnitude of magnetization increases. This may be because of magnetic exchange of $Gd^{3+}$ ions with $Cr^{3+}$ spins. **Table 4** shows the Neel temperature ($T_N$), Spin reorientation temperature ($T_{SR}$), Remanence magnetization ($M_r$) in emu/g and Critical magnetic field ($H_c$) at 10K in kOe of all studied samples as summary of all parameters. Except ECO the Neel temperature ($T_N$) decreases monotonically with the decrease in ionic radii of rare earth ion ($RE^{3+}$). The observed Neel temperature ($T_N$) for NCO is 230.5K and the same is observed 191K for GCO. Due to the exchange interaction between $Cr^{3+}$ spins below the Neel temperature ($T_N$), all studied samples $RECrO_3$ (RE = Nd, Sm, Eu & Gd) shows canted antiferromagnetic transition with weak ferromagnetic ordering at low temperature.

## 4. Conclusion

We conclude that the bulk sample of the reported samples in this paper is synthesized via the solid-state reaction method at the high temperature up to 1400°C. We studied here the Rietveld analysis of the XRD data through which we conclude that all the samples are shown the orthorhombic unit cell structure with space group *Pnma* and *Pbnm*. Calculated Tolerance factor (t) value for all samples lies in between 0.75 to 0.90 that represent the data is standardized data for orthorhombic structure. From the SEM data we conclude that the grain size for reported sample lies between 1.0 to 3.0 μm. Samples NCO, SCO, ECO and GCO shows the Neel temperature ($T_N$) in the range 230K to 177K which is due to the antiferromagnetic ordering of the $Cr^{3+}$ ions. On lowering the temperature, the NCO, SCO and ECO shows spin-reorientation temperature ($T_{SR}$) at 40K, 59K and 24K respectively.

## 5. Acknowledgements

One of us, Anurag Yadav is thankful to Jaypee Institute of Information Technology, Noida for providing financial assistance as Ph.D. fellowship. Authors would like to thank AIRF and SPS at JNU, New Delhi for providing Physical Laboratory for measurement facility (Physical Property Measurement System-14 Tesla). Authors also would like to thank Dr. Abhishek Verma for providing SEM facility at AIARS, Amity University, Noida, Uttar Pradesh. Special thanks to Dr. V. P. S. Awana, Chief Scientist, CSIR-National Physical Laboratory, New Delhi, India for providing his key inputs and fruitful suggestion to improve the manuscript.

**Table 1: Lattice parameters, unit cell volume, χ² value, tolerance factor (t) and orthorhombic strain (S) of all studied samples:**

| Compound Name | | LaCrO$_3$ | NdCrO$_3$ | SmCrO$_3$ | EuCrO$_3$ | GdCrO$_3$ |
|---|---|---|---|---|---|---|
| Lattice Parameter (Å) | *a* | 5.4756(7) | 5.4234(3) | 5.5013(2) | 5.3796(1) | 5.3251(5) |
| | *b* | 7.7648(0) | 5.4807(4) | 7.6451(4) | 5.3957(3) | 5.5082(6) |
| | *c* | 5.5169(8) | 7.6933(5) | 5.3676(1) | 7.6715(0) | 7.6147(9) |
| Cell Volume (Å$^3$) | | 234.569 | 228.667 | 225.750 | 222.702 | 223.360 |
| Space Group | | *Pnma* | *Pnma* | *Pnma* | *Pnma* | *Pbnm* |
| Structure | | Orthorhombic | Orthorhombic | Orthorhombic | Orthorhombic | Orthorhombic |
| χ²value | | 1.54 | 1.49 | 1.63 | 1.86 | 1.56 |
| Tolerance factor (t) | | 0.918 | 0.934 | 0.871 | 0.865 | 0.816 |
| Strain factor (S) | | 0.0075 | 0.3461 | 0.0246 | 0.3512 | 0.3538 |

**Table 2: Rietveld refined atomic coordinates with Wyckoff position of all studied samples:**

| Refined Atomic Co-ordinates with Wyckoff position | | | | | |
|---|---|---|---|---|---|
| Compound Name | Atom | Wyckoff position | *x* (Å) | *y* (Å) | *z* (Å) |
| LaCrO$_3$ (LCO) | La | 4c | 0.0167 | 0.2500 | 0.0037 |
| | Cr | 4a | 0.0000 | 0.0000 | 0.5000 |
| | O1 | 4c | 0.4935 | 0.2500 | 0.0646 |
| | O2 | 8d | 0.2282 | 0.5350 | 0.2285 |
| NdCrO$_3$ (NCO) | Nd | 4a | 0.0000 | 0.0000 | 0.0000 |
| | Cr | 4b | 0.0000 | 0.0000 | 0.5000 |
| | O1 | 4c | 0.5641 | 0.2500 | 0.8745 |
| | O2 | 8d | 0.6542 | 0.4562 | 0.4578 |
| SmCrO$_3$ (SCO) | Sm | 4c | 0.0000 | 0.0000 | 0.0000 |
| | Cr | 4b | 0.0000 | 0.0000 | 0.5000 |
| | O1 | 4c | 0.1453 | 0.0000 | 0.2500 |
| | O2 | 8d | 0.7523 | 0.4682 | 0.6175 |
| EuCrO$_3$ (ECO) | Eu | 1b | 0.5161 | 0.5360 | 0.2467 |
| | Cr | 1a | 0.0000 | 0.5000 | 0.0000 |
| | O1 | 3d | 0.4375 | 1.0276 | 0.2076 |
| | O2 | - | 0.8260 | 1.4990 | 0.0445 |
| GdCrO$_3$ (GCO) | Gd | 4c | 0.9860 | 0.0587 | 0.2500 |
| | Cr | 4b | 0.0000 | 0.5000 | 0.0000 |
| | O1 | 4c | 0.0924 | 0.4782 | 0.2500 |
| | O2 | 8d | 0.6990 | 0.2959 | 0.0440 |

**Table 3: Atomic/weight % of the elements for all samples according to the EDAX data:**

| Sample Name | Elements | Weight % | Atomic % |
|---|---|---|---|
| | C K | 1.12 | 5.66 |
| **LaCrO$_3$** | O K | 9.31 | 35.43 |
| | La L | 62.79 | 27.53 |
| | Cr K | 26.79 | 31.38 |
| **NdCrO$_3$** | C K | 2.15 | 9.29 |
| | O K | 13.01 | 42.22 |
| | Nd L | 56.75 | 20.43 |
| | Cr K | 28.10 | 28.06 |
| **SmCrO$_3$** | C K | 0.59 | 3.72 |
| | O K | 6.00 | 28.32 |
| | Cr K | 22.11 | 32.13 |
| | Sm L | 71.30 | 35.84 |
| **EuCrO$_3$** | O K | 12.30 | 47.01 |
| | Cr K | 22.88 | 26.90 |
| | Eu L | 64.82 | 26.08 |
| **GdCrO$_3$** | O K | 10.49 | 42.69 |
| | Cr K | 24.17 | 30.27 |
| | Gd L | 65.34 | 27.05 |

**Table 4: Values of Neel Temperature ($T_N$), Spin reorientation ($T_{SR}$), Remanence Magnetization ($M_r$) and Critical Field ($H_c$) for studied samples:**

| Sample Name | $T_N$ (K) | $T_{SR}$ (K) | $M_r$ (emu/g) at 10K | $H_c$ (kOe) at 10K |
|---|---|---|---|---|
| NdCrO$_3$ | **230.5** | **40** | 0.1209 | 0.959 |
| SmCrO$_3$ | **201.0** | **59** | 0.0415 | 1.796 |
| EuCrO$_3$ | **177** | **24** | 1.756 | 0.089 |
| GdCrO$_3$ | **191** | **-** | 0.73 | 10.25 |

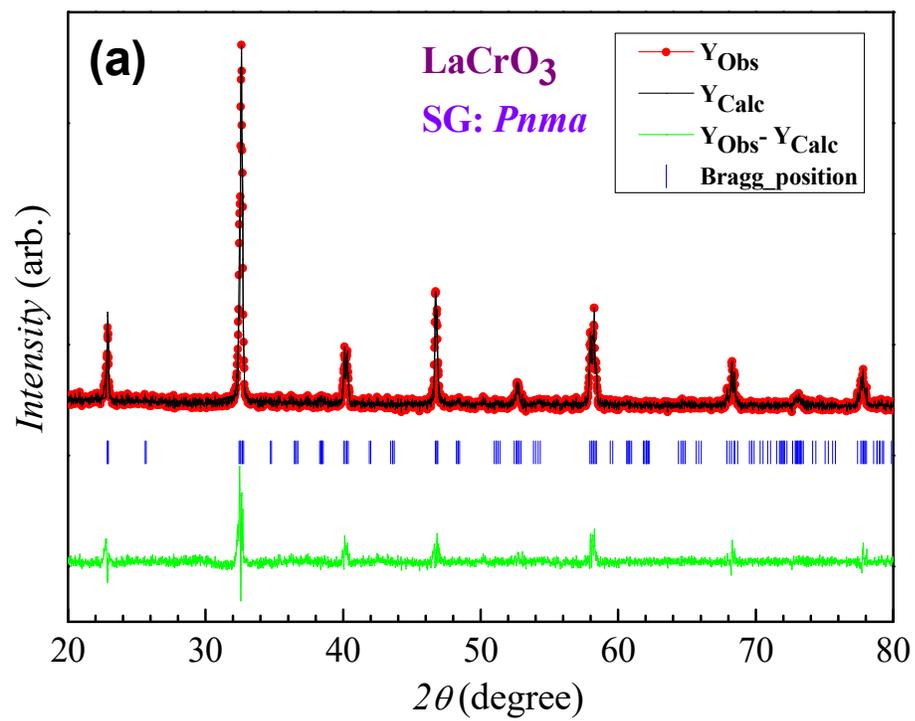
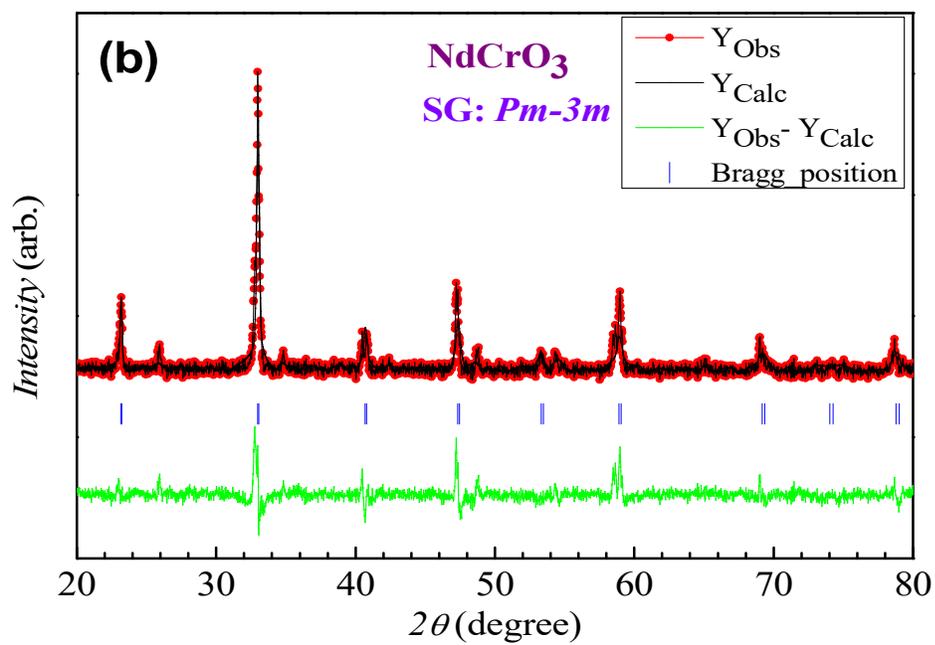

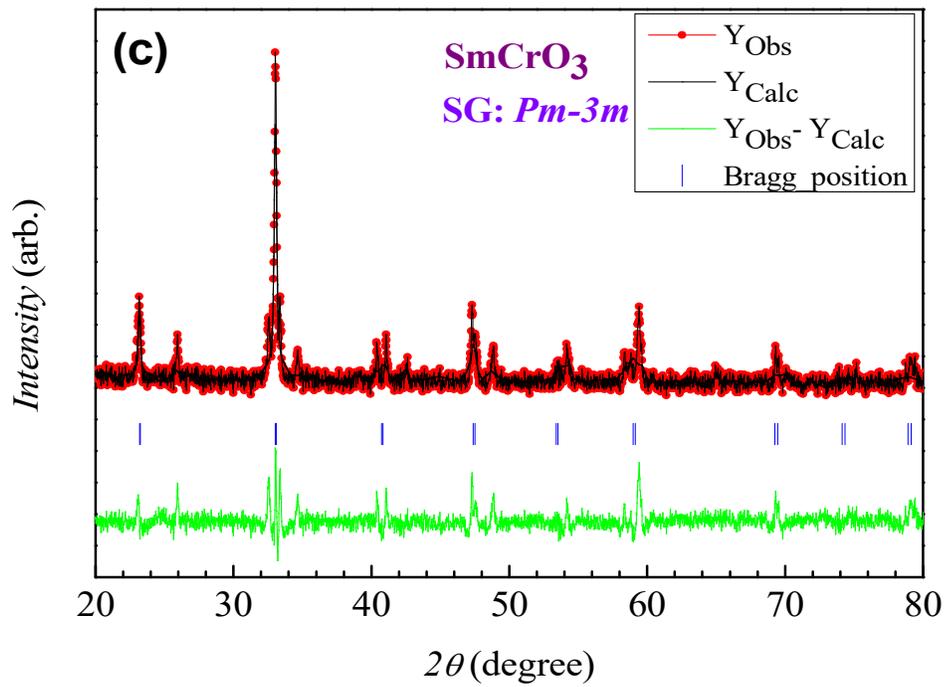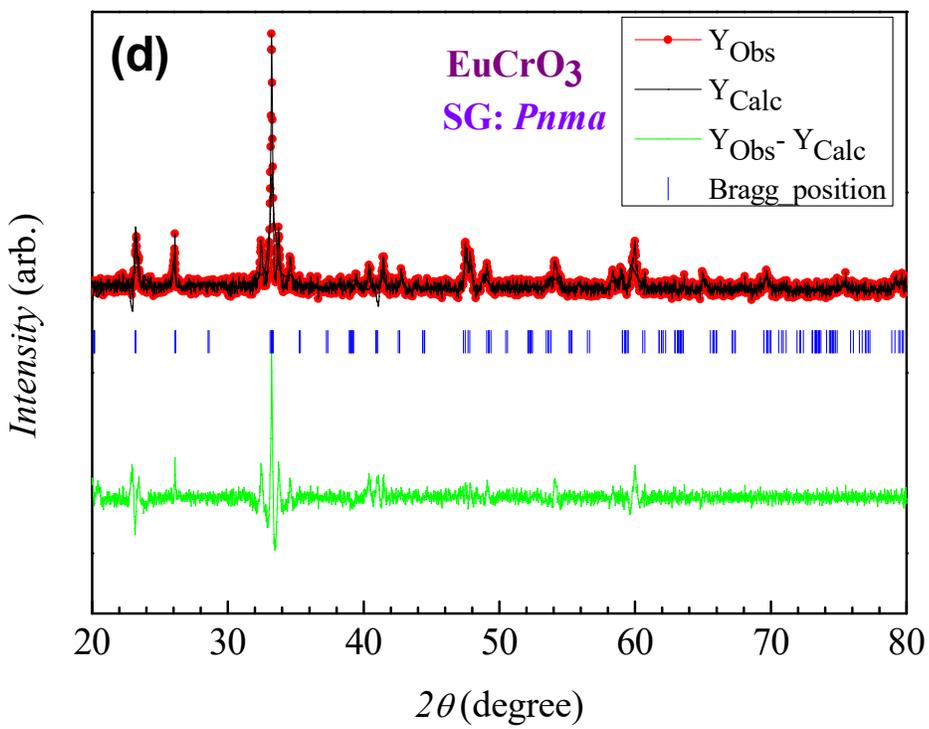

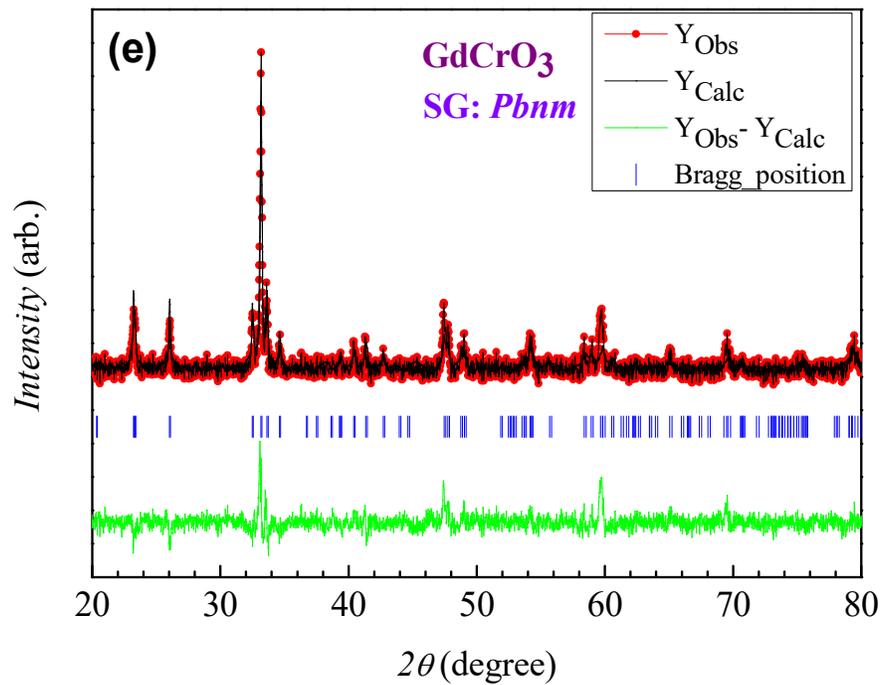

**Figure 1:** Observed and calculated XRD patterns of **(a)** LaCrO$_3$, **(b)** NdCrO$_3$, **(c)** SmCrO$_3$, **(d)** EuCrO$_3$, and **(e)** GdCrO$_3$. Solid lines at the bottom are the difference between the observed and calculated patterns. Vertical solid lines at the bottom show the position of allowed Bragg peaks.

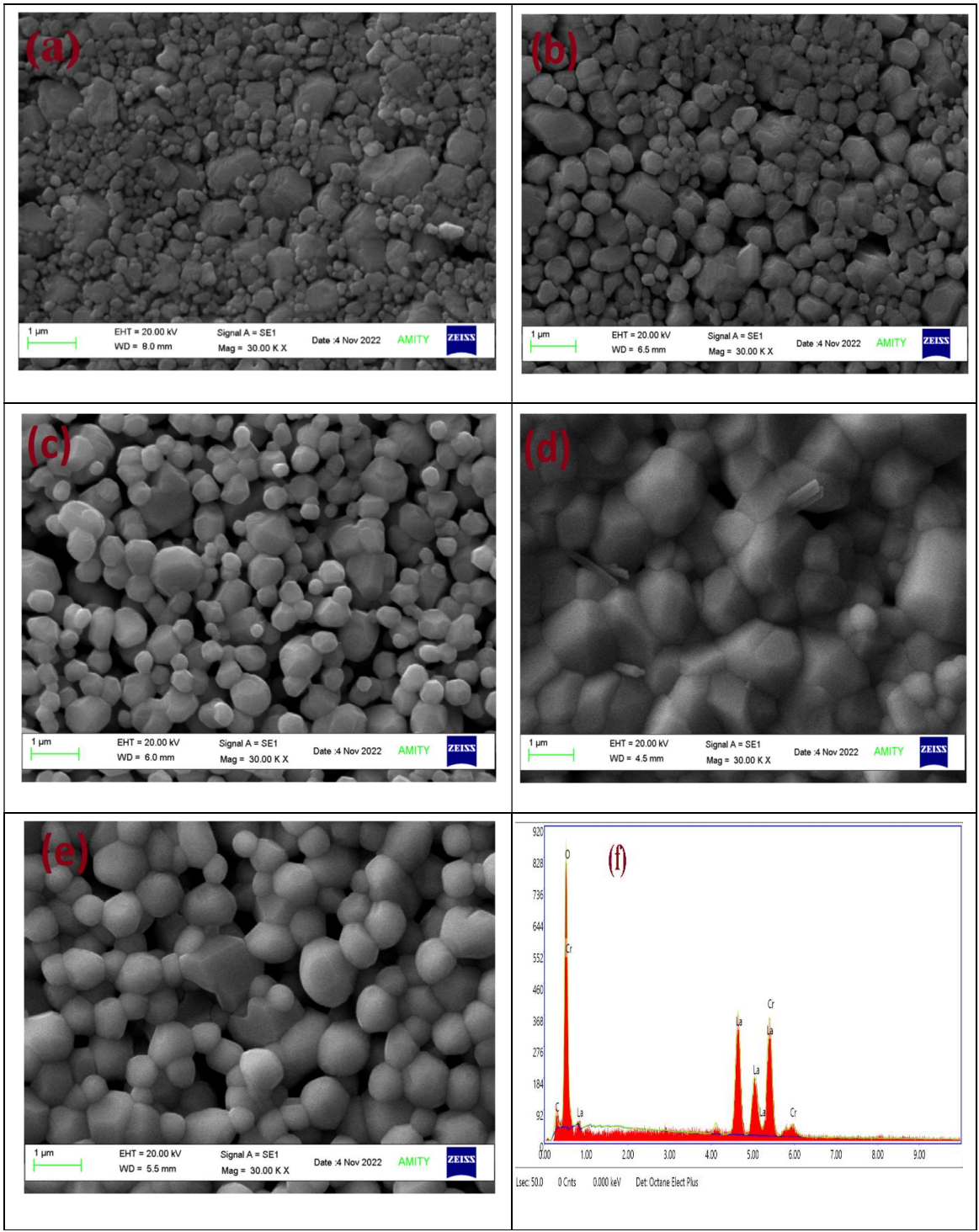

**Figure 2:** SEM images of the sintered pellets, for **(a)** LaCrO$_3$, **(b)** NdCrO$_3$, **(c)** SmCrO$_3$, **(d)** EuCrO$_3$, and **(e)** GdCrO$_3$ at 30K magnification and EDAX spectrum for **(f)** LCO.

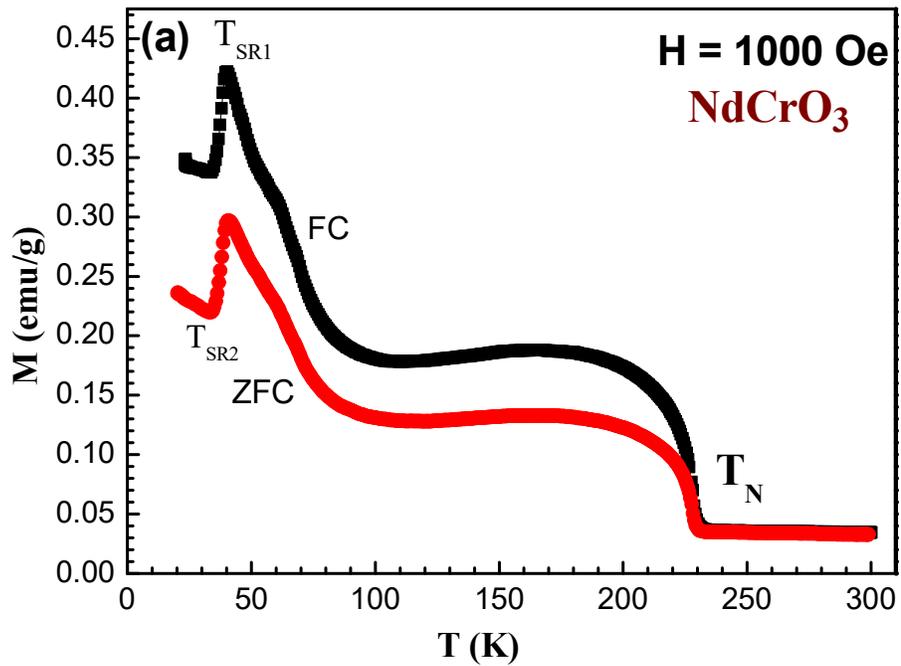
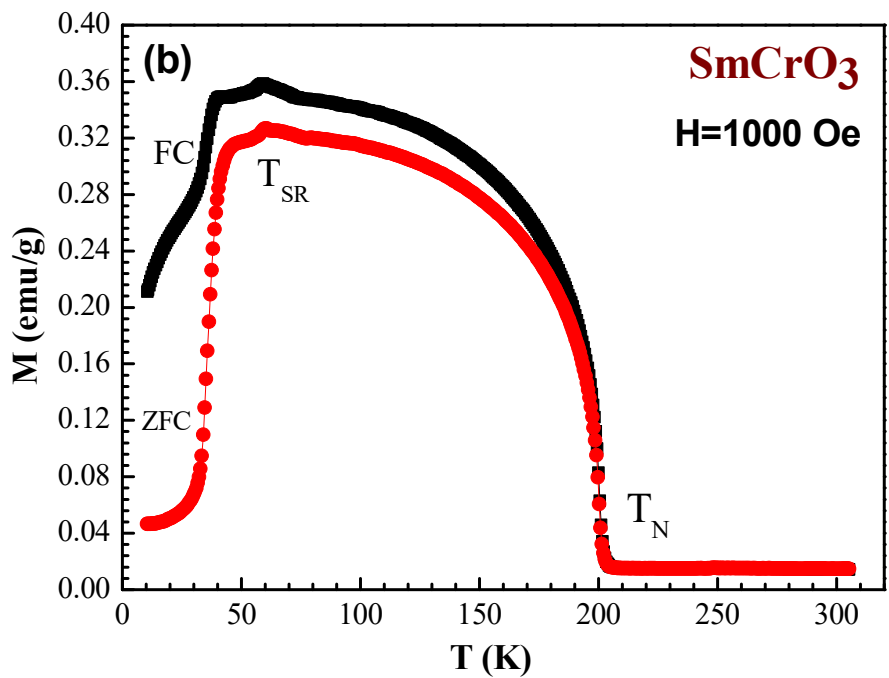

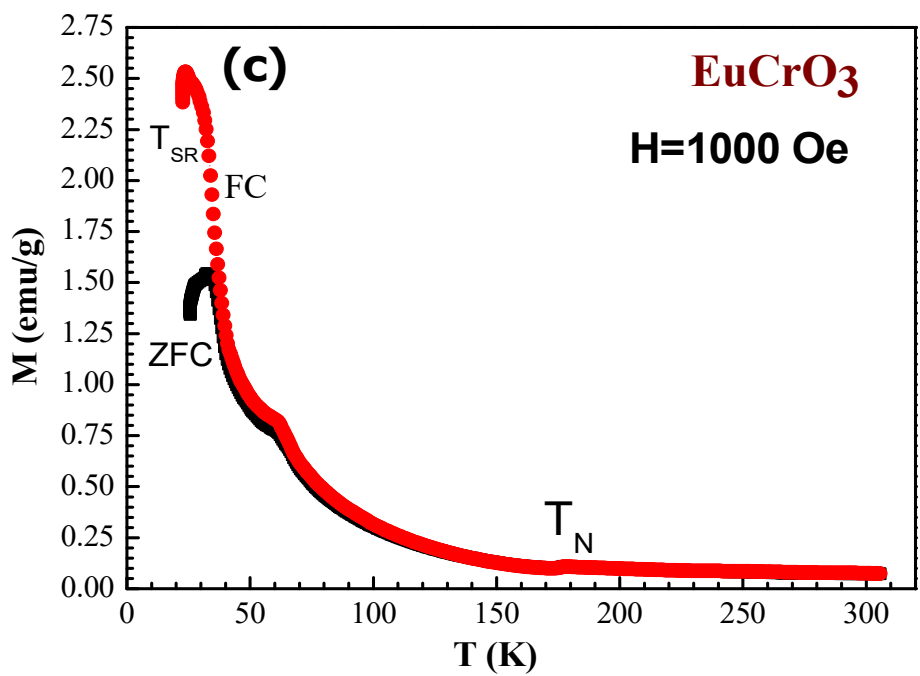

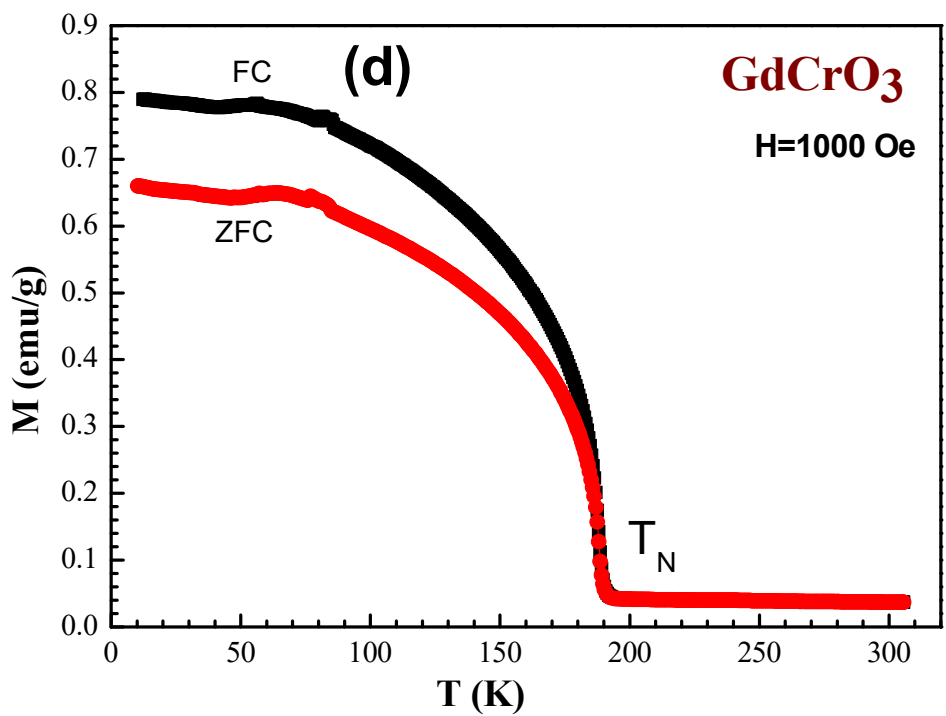

**Figure 3:** ZFC and FC dc magnetization plots for **(a)** NdCrO$_3$, **(b)** SmCrO$_3$, **(c)** EuCrO$_3$, and **(d)** GdCrO$_3$ measured in applied field H = 1000 Oe.

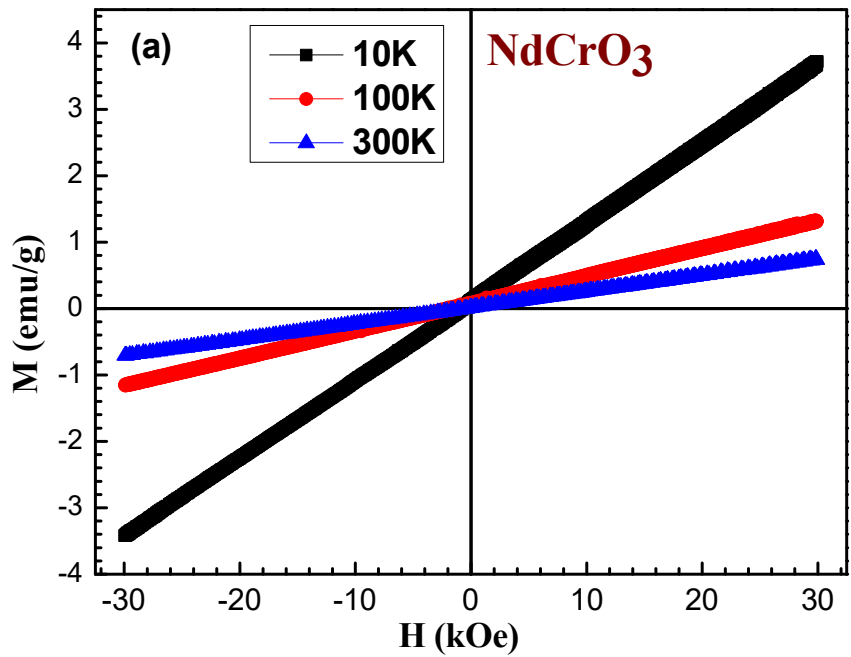

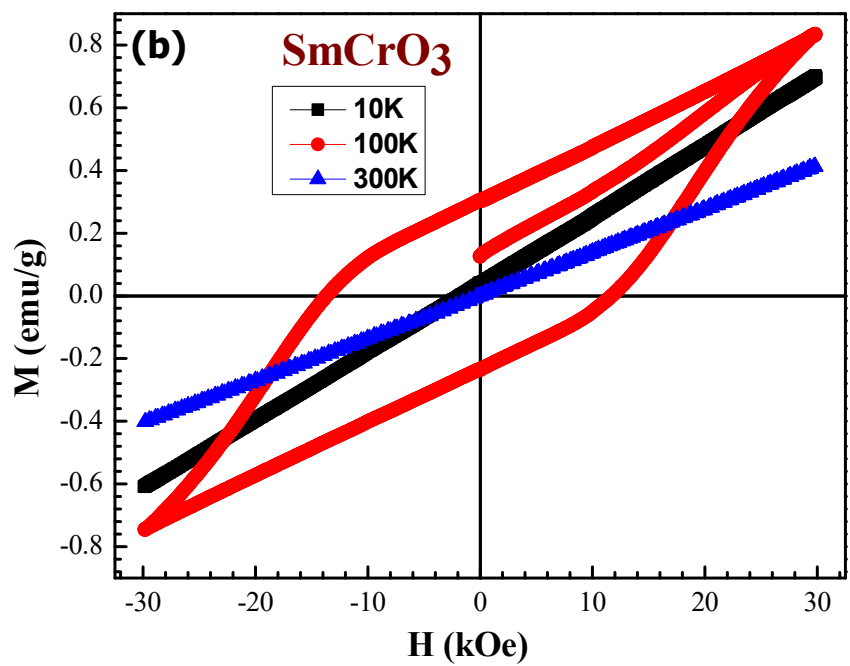

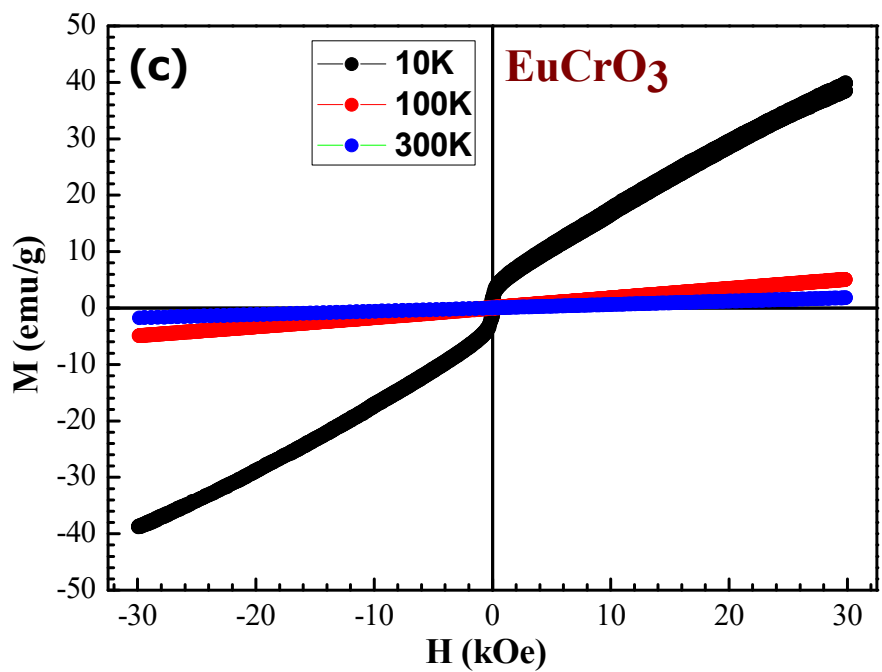

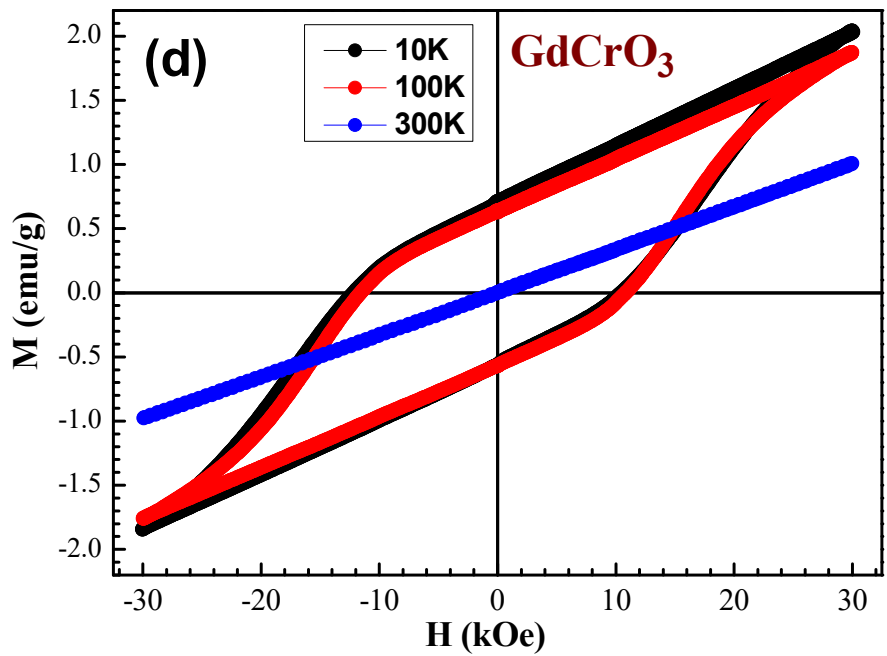

**Figure 4:** DC magnetization as a function of applied magnetic fieldfor **(a)** NdCrO$_3$, **(b)** SmCrO$_3$, **(c)** EuCrO$_3$, and **(d)** GdCrO$_3$ measured at different temperatures (5, 100, and 300K) in the range - 30kOe to + 30kOe.